# Multi-Channel Feature Extraction for Virtual Histological Staining of Photon Absorption Remote Sensing Images

Marian Boktor[1,2], James E.D. Tweel[1,3], Benjamin R. Ecclestone[1,3], Jennifer Ai Ye[2], Paul Fieguth[2], Parsin Haji Reza[1,*]

[1]PhotoMedicine Labs, University of Waterloo, 200 University Ave W, Waterloo, ON N2L 3G1
[2]Vision and Image Processing Lab, University of Waterloo, 200 University Ave W, Waterloo, ON N2L 3G1
[3]illumiSonics Inc., 22 King Street South, Suite 300, Waterloo, ON, N2J 1N8
*Corresponding author: phajireza@uwaterloo.ca

**Abstract**

Accurate and fast histological staining is crucial in histopathology, impacting diagnostic precision and reliability. Traditional staining methods are time-consuming and subjective, causing delays in diagnosis. Digital pathology plays a vital role in advancing and optimizing histology processes to improve efficiency and reduce turnaround times. This study introduces a novel deep learning-based framework for virtual histological staining using photon absorption remote sensing (PARS) images. By extracting features from PARS time-resolved signals using a variant of the K-means method, valuable multi-modal information is captured. The proposed multi-channel cycleGAN (MC-GAN) model expands on the traditional cycleGAN framework, allowing the inclusion of additional features. Experimental results reveal that specific combinations of features outperform the conventional channels by improving the labeling of tissue structures prior to model training. Applied to human skin and mouse brain tissue, the results underscore the significance of choosing the optimal combination of features, as it reveals a substantial visual and quantitative concurrence between the virtually stained and the gold standard chemically stained hematoxylin and eosin (H&E) images, surpassing the performance of other feature combinations. Accurate virtual staining is valuable for reliable diagnostic information, aiding pathologists in disease classification, grading, and treatment planning. This study aims to advance label-free histological imaging and opens doors for intraoperative microscopy applications.

## I. INTRODUCTION

Histopathology is a branch of pathology that utilizes microscopic examination of chemically stained thin tissue sections to investigate and diagnose diseases by analyzing cellular and structural abnormalities. Traditional histology has several limitations including time-consuming procedures, tissue alteration, limited staining options, and variability [1], [2]. Consequently, there is a growing need for imaging techniques that can directly visualize cells and tissues in their native states without the need for chemical labeling. As such, label-free microscopy techniques have gained significant attention in histopathology as they offer non-invasive methods to visualize cells and tissues [3]–[7] through leveraging the intrinsic properties of biological samples.

Label-free techniques acquire informative contrast from unstained samples while preserving their integrity for subsequent analysis. These label-free contrasts can be extracted and transformed into familiar formats, such as conventional histopathology stains, through the process of virtual staining. Label-free contrasts from various optical microscopes have been coupled with deep learning algorithms to intelligently present the rich information in meaningful ways. Examples of such microscopes include quantitative phase imaging [8], reflectance confocal microscopy [3], photoacoustic microscopy [9], [10], and autofluorescence microscopy [7]. The specificity level to biomolecules varies for each microscope, and as the microscope becomes more specific towards biomolecules of interest, the dependence on the deep learning model for inference decreases. For instance, autofluorescence shows promise in capturing extranuclear elements, however, important structures like cell nuclei do not exhibit measurable autofluorescence [11]. The lack of direct measurement of nuclear contrast may cause deep learning to estimate the coloring of nuclear structures.

Autofluorescence is a form of radiative relaxation and re-emittance of lower energy photons following the absorption of light. However, a portion of the absorbed energy can also undergo relaxation via temperature and pressure. These are termed non-radiative relaxation pathways and they provide additional contrast and hold significant value in capturing critical structures, including cell nuclei (DNA) [12]. One modality, Photon Absorption Remote Sensing (PARS) microscopy, previously known as Total-Absorption Photoacoustic Remote Sensing [12], [13], can simultaneously measure the radiative (e.g. autofluorescence) and non-radiative (temperature and pressure) relaxation of biomolecules. By leveraging the unique absorption spectra of biomolecules like hemoglobin, DNA, and lipids, PARS enables direct selective imaging without the need for extensive systematic modifications [14]–[16]. PARS has emerged as a powerful method that utilizes a non-contact, all-optical architecture to visualize the intrinsic absorption contrast of biological tissue structures.

In PARS, a pulse of light is used to excite the sample and induce both radiative and non- radiative relaxations processes which are then measured [13]. The optical emissions generated from the radiative relaxation can be directly measured while the non- radiative relaxation results in nanosecond-scale variations in the specimen's optical properties. These non-radiative modulations can be measured as forward or backscattered intensity fluctuations using a secondary co-focused detection laser [13]. Furthermore, optical scattering contrast is also measured using the pre-excitation scattering intensity of the detection laser. All three contrasts are simultaneously measured from a single excitation event and are therefore intrinsically co-registered.

Previously, the PARS microscope has demonstrated its efficacy in generating virtual H&E stains using a pix2pix image-to-image translation model [10]. However, pix2pix demands accurate alignment of PARS and true H&E images at the pixel level. Pixel-level registration is not always achievable due to alterations in the tissue during the staining process. Poorly aligned pairs weaken the colorization performance and lead to distorted or blurred translations [17]. To avoid such problem, here we choose to employ a more flexible image-to-image translation model known as cycleGAN, short for Cycle-Consistent Generative Adversarial Network [18]. While cycleGAN can be trained with unpaired data, in this study, registered pairs are used as they help enhance the colorization results achieved by cycleGAN [17], primarily through mitigating hallucination artifacts.

Previous PARS virtual staining efforts used both the non-radiative and radiative channels from a single 266 nm UV excitation wavelength [10], [17], [19]. While these combined contrasts are highly analogous to the nuclear and connective tissue contrasts highlighted with H&E, some structures like red blood cells are not strongly captured with UV excitation alone. An additional excitation wavelength can be used to target hemoglobin or red blood cells to provide better separation between them and comparably sized features, such as nuclei. Providing additional information to the model may help it understand the separability between different structures and enhance its ability to learn the statistical relationship between PARS and H&E domains.

Accordingly, an additional 532 nm excitation laser is employed in this study to help acquire red blood cell contrast, resulting in the non-radiative time-domain (TD) signals exhibiting two excitation peaks. The first peak, at a wavelength of 266 nm, is used to specifically target DNA/RNA, while the second peak, at 532 nm, primarily targets hemoglobin (red blood cells). Moreover, recently Pellegrino et al. [20], [21] showed that multiple features may be extracted from a single peak time-resolved non-radiative relaxation signal. These features are extracted based on signal shape and may relate to more specific tissue structures [20], [21]. Hence, we choose to expand the number of input channels to the cycleGAN model through intelligently extracting features from the non-radiative TD signals to improve the distinction between different structures. To extract such features, we utilize the K-means method presented in [20]. These features are subsequently employed to reconstruct feature images, which together with the conventional non-radiative and radiative images form an array of image components. Such components can serve as inputs for the virtual staining cycleGAN model in different arrangements. The novelty of this work resides in the use of such features to train a cycleGAN model.

CycleGAN conventionally allows single- or three-channel input data. To effectively utilize the extracted features in virtual staining, this study introduces multi-channel cycleGAN (MC-GAN) that extends the existing cycleGAN model used in previous work [17] to support more than three channels. This approach allows for the extraction of multiple features, providing a better understanding of the potential of the TD signals in improving virtual staining. The key contribution of this work is to enhance the performance of the colorization model by incorporating multiple features for labeling structures. Another contribution of this work is the introduction of a comprehensive pipeline that enables the extraction and selection of the most effective features to enhance the process of colorization.

## II. METHODS

### A. Sample Preparation and Data Acquisition

The study utilized a dataset obtained from thin sections of formalin fixed paraffin embedded (FFPE) human skin tissues and mouse brain tissues. The human tissue samples are provided by the Alberta Precision Laboratories in Calgary, Alberta, Canada. The collection of these samples adhered to approved protocols established with the Research Ethics Board of Alberta (Protocol ID: HREBA.CC-18-0277) and the University of Waterloo Health Research Ethics Committee (Protocol ID: 40275). The requirement for patient consent was waived by the ethics committee, as samples are anonymized to remove all patient information, and tissues are archival samples not required for patient diagnostics. All experiments involving human tissues are conducted in compliance with the relevant guidelines and regulations of the government of Canada, such as "Ethical Conduct for Research Involving Humans (TCPS 2)". The mouse brain samples were prepared at the National Institute of Health in Bethesda, Maryland, United States. These samples adhered to approved protocols under the University of Waterloo Health Research Ethics Committee (Protocol ID: 44595).

The preparation of unstained tissue sections involves several steps. Initially, tissue is resected and immediately fixed in a 10% neutral buffered fixative for up to 48 hours. Next, the sample is submerged in a series of alcohol exchanges to dehydrate the tissue. Dehydration is followed by a series of rinses in xylene, a tissue clearing agent which removes any fat residues. The dehydrated and cleared tissue is then soaked in molten paraffin to allow for the complete infiltration of paraffin into the tissue. Finally, tissues are embedded in paraffin and allowed to solidify at room temperature, forming a FFPE tissue block. Tissue sections are prepared by using a microtome to slice thin sections of about 3-5 µm from the FFPE block. Thin sections are then transferred to a glass microscope slide by means of a water bath and allowed to dry. Slides are heated to 60 °C for 60 minutes in a laboratory to evaporate excess paraffin. After this process, the thin sections are ready for imaging using the PARS system. Once the PARS dataset is acquired, the thin tissue slides are stained using H&E. The stained slides are completed with a mounting medium and a coverslip. The stained sections are imaged using a transmission mode brightfield microscope, thereby generating the corresponding H&E ground truth dataset.

To create datasets for training the virtual staining GAN model, a collection of PARS images and corresponding H&E images is prepared as follows. Thin, unstained tissue sections are scanned using the PARS system, as outlined in Section II.B. After imaging, specimens are stained with H&E, then scanned using a brightfield microscope. This results in matched pairs of PARS and H&E images from the same samples. In Figure 1, a comparison between the conventional histochemical staining process and our proposed virtual staining process is presented. For training the virtual staining model, samples first undergo the PARS imaging pathway, then the standard staining procedure. This is necessary for generating ground truth data. However, this step is not required once the model has been trained.

## B. PARS Imaging

While this paper does not focus on the PARS system design, it is important for understanding the data collection process. For a more detailed exploration of the system architecture and image formation process, refer to "Automated Whole Slide Imaging for Label-Free Histology using Photon Absorption Remote Sensing Microscopy" by Tweel *et al.* [19]. Briefly, the experimental setup is depicted in Figure 2. This architecture features two excitation sources: a 266 nm UV laser (Wedge XF, RPMC) and a 532 nm visible laser (Wedge XF, RPMC). The detection source is a 405 nm OBIS-LS laser (OBIS LS 405, Coherent). The excitation and detection sources are combined using a dichroic mirror and focused onto the sample using a 0.42 NA UV objective lens (NPAL-50-UV-YSTF, OptoSigma). This configuration provides a maximum spatial resolution of approximately 400 nm. The detection light, containing the optical scattering and non-radiative relaxation is collected after transmission through the sample using a second objective lens (100X

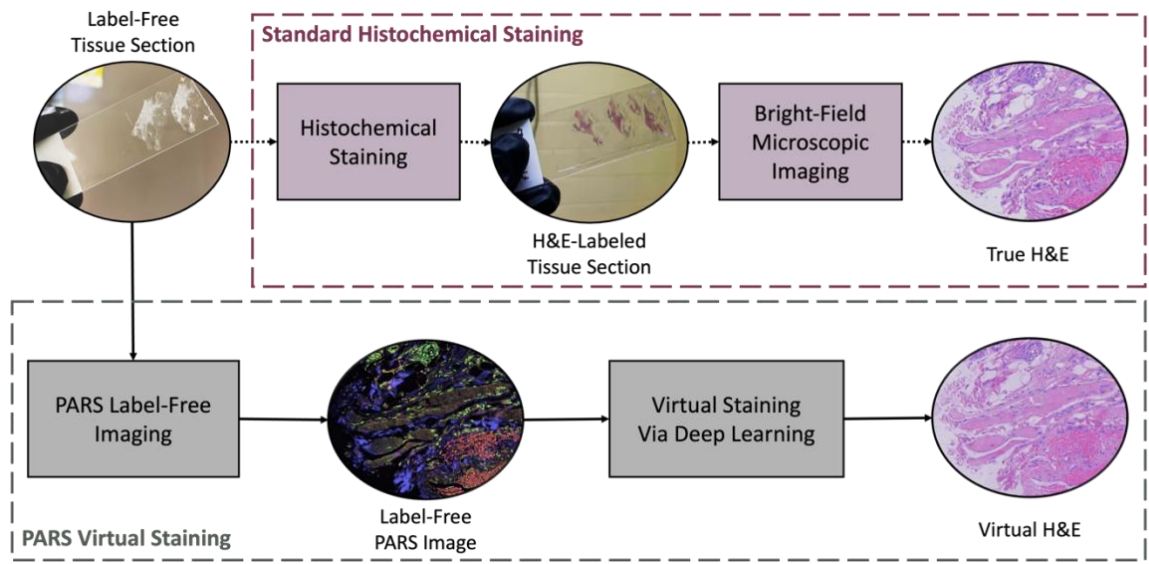

**Figure 1.** Virtual Staining of PARS images via deep learning. The top pipeline shows the standard workflow to generate images of histochemical stains. The bottom pipeline shows the steps for virtual staining. The same tissue section is used in the two workflows for the deep model training and performance analysis.

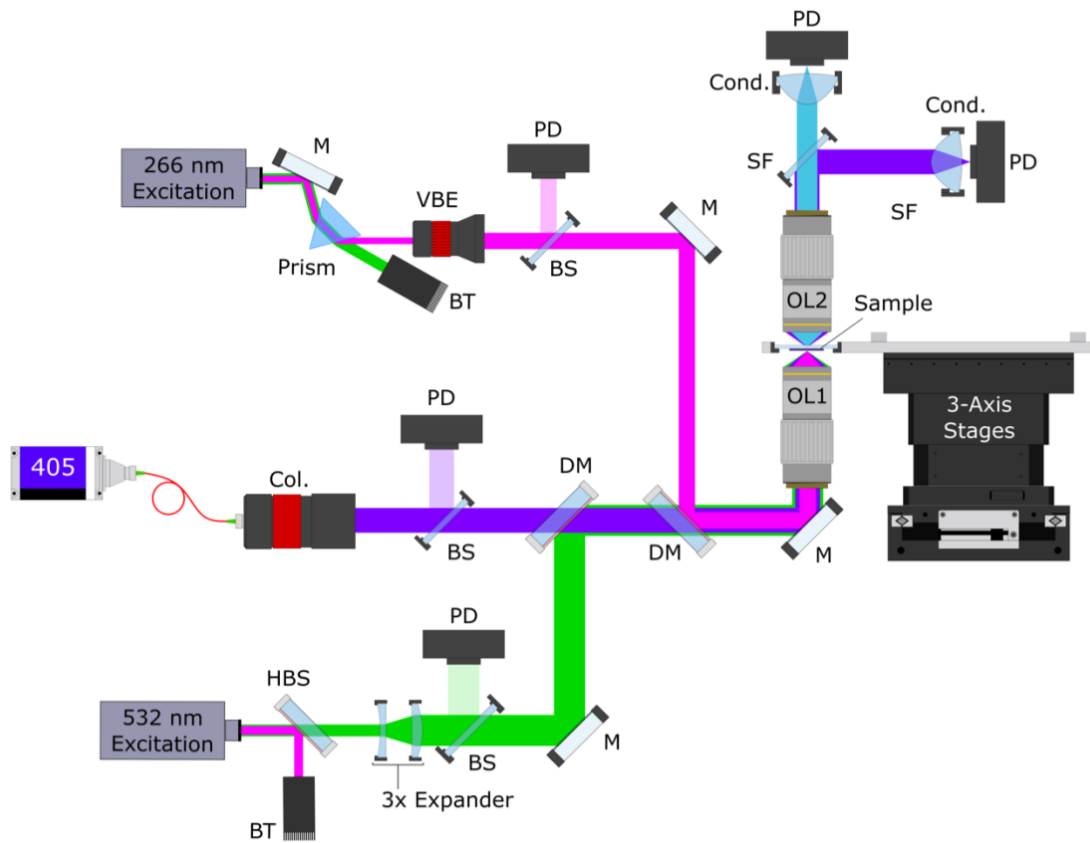

**Figure 2.** Simplified PARS histology optical architecture. Component labels are defined as follows: mirror, *M*; dichroic mirror, *DM*; variable beam expander, *VBE*; collimator, *Col.*; condenser lens, *Cond.*; spectral filter, *SF*; beam sampler, *BS*; photodiode, PD; objective lens, OL; harmonic beam splitter, *HBS*; beam trap, *BT*.

Mitutoyo Plan Apo, Mitutoyo). The same lens also collects the radiative emissions from the sample. To separate the non-radiative detection light from the radiative relaxation, a spectral filter (NF405-13, Thorlabs) is employed. Then, the radiative emission amplitudes are directly measured using a photodiode (APD130A2, Thorlabs). Concurrently, the non-radiative detection light is directed to another photodiode (APD130A2, Thorlabs) where the optical scattering and the absorption-induced intensity modulations are captured.

PARS images are generated through the following process. The sample is scanned over the stationary objective lens using mechanical stages. The 266 nm and 532 nm excitation sources are pulsed continuously at a rate of 50 kHz, with energies of ~150-200 pJ, and ~2.5 nJ for the 266 nm and 532 nm source, respectively. Concurrently, the continuous wave detection laser operates at ~3-5 $\mu$W during imaging. The 532 nm pulses are synchronized to occur ~500 ns after the 266 nm pulses. The stage motion is then tuned to ensure the 266 nm excitation event occurs every 250 nm. During each excitation event, the optical scattering, and radiative and non-radiative relaxation signals are collected from the corresponding photodiodes, for each excitation wavelength. Additionally, a position signal is recorded from the scanning stages. The radiative signals are condensed into a single feature by extracting the amplitude of the measured signal. The optical scattering is extracted by averaging the transmitted detection prior to excitation. The post-excitation time-resolved non-radiative relaxation modulations are stored in their entirety for further processing.

To generate a baseline model for comparisons, the non-radiative signal reconstruction method used in previous embodiments is employed [10]. This method involves signal extraction by integrating the post excitation modulation energy of the non-radiative signals. The baseline is used to contrast the efficacy of intelligently extracting features (explained in Section II.D) from the non-radiative signals for enhancing the virtual staining process.

### C. Image Registration and Preprocessing

In this study, we choose to use registered PARS and H&E pairs for training a cycleGAN for virtual staining, as discussed in Section I. However, the PARS and H&E images are not inherently co-registered. This is attributed to the H&E staining procedure, and the two different acquisition processes. Subsequently, additional steps, including field-of-view matching and registration, are required to align the PARS and H&E images properly for model training and performance analysis.

The preprocessing and registration process follows the workflow previously described by Boktor et al. [10]. This includes extracting small fields from the PARS and H&E images, then coarsely matching them for one-to-one registration. The control point registration tool is used from the MATLAB Image Processing Toolbox is used for registration, with PARS as the reference images and H&E as the registered images. Control points are manually selected and refined to minimize distortions. A non-rigid geometric transformation [22] is then fitted between the two images, and the transformation is applied to the H&E images, resulting in co-registered pairs of PARS and H&E images. Following registration, PARS images are preprocessed as follows. Each image is normalized, then the contrast is enhanced by saturating the highest and lowest 1% of pixel values. Finally, a color reversal is performed to match the PARS images with the grayscale ground truth images. The same preprocessing is applied to all datasets used in this study.

### D. Time-Domain Feature Extraction

To extract material-specific information from non-radiative TD signals, a method is needed to identify constituent time-domain features that accurately represent the underlying tissue target. The method, proposed by Pellegrino et al. [20], is utilized here which features K-means clustering with a modified approach to compute cluster centroids. To generate feature images from the non-radiative TD signals, two steps are involved feature learning and feature extraction.

The intelligent clustering aims to identify $K$ characteristic shapes in the signals, described as a set of $K$ centroids, $\mathcal{F} = \{f_i(t)\}, i = 1, \ldots, K$. TD signals are treated as Cartesian vectors in space $\mathbb{R}^n$, where $n$ corresponds to the number of TD samples, and thus the shape of the signal is associated with the angle of the corresponding vector, and the distance between TD signals is quantified by the sine of the angle between them, resulting in a maximum distance for orthogonal signals and zero distance for scaled or inverted signals. Cluster centroids are computed as the principal component of the combined set of each cluster and its negative, ensuring that the learned centroids are resilient to noise.

Following the K-means clustering, a set of feature vectors $\mathcal{F} = \{\vec{f}_i\}$ is obtained, which can represent the signals as a weighted sum. These feature vectors are then arranged in the form of a matrix of features, $F = [\vec{f}_1 | \vec{f}_2 | \ldots | \vec{f}_K]$. The amplitudes of the learned TD features (centroids) contained within each time domain are extracted by transforming from the time-domain to the feature-domain. This is performed by multiplying each TD signal with the pseudo-inverse of $F$ [20], [23]. The result is an array of $K$ feature images, $M_f = [m_{f_1}, m_{f_2}, \ldots, m_{f_K}]$.

The appropriate value of $K$ is determined according to the specific dataset as discussed in Section III. A minimum of 2 clusters is examined during the feature extraction process. However, it is important to impose an upper limit on $K$, set to 6 in this study, to prevent the generation of redundant clusters and avoid the introduction of visually indistinguishable or uninformative features. Following the determination of the optimal value for $K$, a set of $K$ feature images is generated. These feature images alongside the conventional non-radiative and radiative image components form an array of images, which can serve as the inputs for the colorization model in different combinations.

### E. Multi-Channel GAN (MC-GAN)

This study applies cycleGAN to convert label-free PARS images into virtually stained images that resemble their corresponding H&E histochemical stained samples. CycleGAN learns feature relation between an input image domain, $A$ and a target image domain $B$, and generates the generators $G: A \to B$ and $F: B \to A$ and the discriminators $D_A$ and $D_B$. The generators aim to generate realistic images that resemble the target domain while preserving the essential characteristics of the input domain. The discriminator, on the other hand, tries to differentiate between real images from the target domain and fake images produced by the generator. It provides feedback to the generator by assessing the fidelity of the generated images, enabling the generator to improve its output quality over time through an adversarial training process.

In this work, we assume domain *A* is PARS imagery while domain *B* is H&E imagery. CycleGAN is able to operate in a scenario where paired examples are not available, nevertheless the absence of paired samples poses challenges in mapping between the source and target domains as it is an under-constrained. To tackle this, CycleGAN incorporates an inverse mapping and introduces a cycle consistency loss [18]. This loss enforces that the translated images can be reliably reversed to their original form. However, it is worth noting that although paired examples are not required for training, the utilization of paired samples in our CycleGAN training does lead to visible performance improvements [17].

Typically, cycleGAN models are trained using single- (grayscale) or three-channel (RGB) images [9], [18], [24]. In previous works [10], [17], [19], these RGB channels were directly replaced with the non-radiative and radiative channels, which posed no issues as the number of channels did not exceed three. However, the focus of this research is to improve the efficacy of the virtual staining process using intelligently extracted features from non-radiative PARS signals, resulting in a scenario where the colorization model has more than three input channels. It is important to note that the use of three channels in conventional cycleGAN models is solely due to the nature of working with RGB images, and there is no inherent reason to restrict the input channels to three in principle-components or data-fusion methods. In fact, the optimal number of input channels is likely to differ from three. Hence, the proposed MC-GAN model enables the incorporation of additional features by expanding the number of input channels beyond three. The integration of extra channels in the MC-GAN model allows for a wider range of information to be utilized during training. This extension enhances the model's ability to capture and leverage diverse information, which can potentially lead to improved colorization performance.

In cycleGANs, *N*-channel input generates *N*-channel output. However, our target H&E domain is an RGB image with three channels. Only a three-channel output is required, regardless of the number of channels in the source domain. To allow a three-channel output with an *N*-channel input, we duplicate the last channel (B) in the target domain to expand the target H&E domain dimensions to match the source PARS domain. When extracting the colorization results, we discard these duplicated channels. Except for the modifications of the multi-channel input and output, the architecture utilized in this study remains identical to the original cycleGAN.

### *F. Training Settings*

Two datasets are employed in this work: human skin and mouse brain data. To generate the training sets for the two datasets, overlapping patches of 256 × 256 pixels are extracted from the PARS and H&E images. For the human skin dataset, approximately 500 overlapping patches are extracted, while for the mouse brain dataset, around 2000 overlapping patches are extracted. The pairs are split in a ratio of 70% for training and

30% for validation. The learning rate is set to 0.0002. The maximum number of epochs is set to 200 with an early stopping criterion to terminate the training when the generator loss stops improving. The trained model is then applied to the test images which are also subdivided into overlapping patches of 256 × 256 pixels. An overlap of ~ 50% is usually sufficient to avoid visible artifacts at the borders of adjacent patches in the final stitched image. The colorization algorithm is implemented in Python version 3.10.6 and model training is implemented using PyTorch version 2.0.0 with support of CUDA version 12.

### III. RESULTS AND DISCUSSION

In previous PARS virtual staining embodiments [10], [17], [19], only the non-radiative and radiative (denoted as NR and R, respectively, in this section for notation simplicity) channels are used as inputs for the virtual staining model. Traditionally, NR images are created by integrating the post excitation modulations in the detection signal. This method omits valuable temporal information because the shape of the signal may contain information associated with specific biological structures [20], [21]. In this paper, we opt to broaden the range of input channels by incorporating features extracted from the NR TD signals. These features augment information at each pixel location, which can enhance the colorization model's ability to comprehend the statistical transformation between the input and the target domains.

The multi-channel virtual staining workflow is shown in Fig. 3. The pipeline consists of two main parts: (1) feature learning, extraction, and selection, and (2) MC-GAN model training. Feature learning, of $K$ features, takes place using a representative subset of the NR TD signals. Feature extraction is performed on all the TD signals to form $K$ feature images. These $K$ different feature images, along with the NR signal integral (as described in Section II.B) from each excitation wavelength (266 nm and 532 nm), and the R channel from 266 nm excitation only, are then fed into the feature selection stage. Feature selection is used to enhance the model's prediction power by eliminating redundant data, increasing contrast between the selected features, and reducing the training volumes and times. Finally, the images of the selected features are used as an input to the proposed MC-GAN model, and the true H&E serves as the ground truth.

A study (the "$K$-study") was conducted to determine the optimal number of features $K$ to extract from each section. In this $K$-study, feature extraction is performed using the K-means algorithm (presented in Section II.D) generating $M_f^K$ for every $K \in \{2, ... , 6\}$. Then, the MC-GAN is trained using the R channel and $M_f^K$ for each $K$. Since the $K$ features are extracted from the NR channel TD signals, they are independent of the R contrast. Hence, the R channel is always used as an additional component in the virtual staining phase of the $K$-study for a fair comparison. The model performance is then assessed.

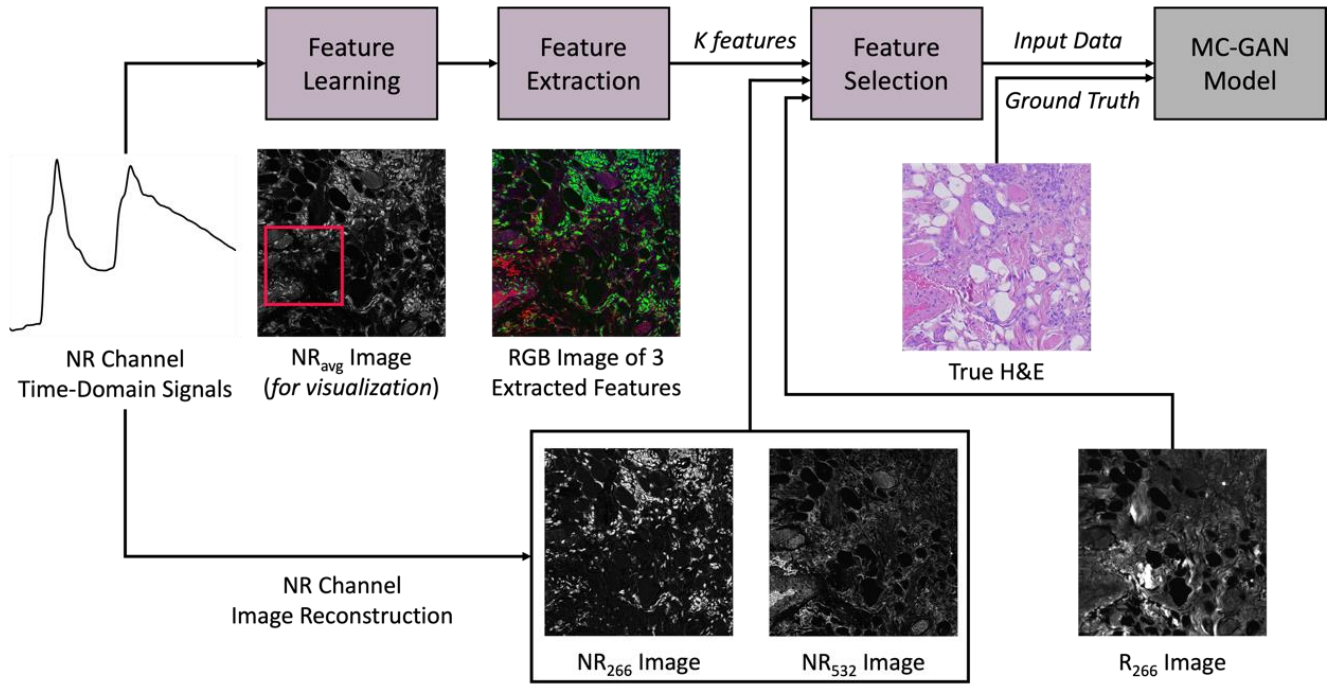

**Figure 3.** Multi-Channel Virtual Staining Workflow. First, feature learning, of $K$ features, takes place using a subset (shown in red box) of the NR channel TD signals. Second, feature extraction is performed on all the TD signals of the data in hand forming $K$ feature images. NR images of each excitation wavelength (266 nm and 532 nm in this case) and R images are extracted separately and passed along with the $K$ feature images to the feature selection phase. The selected features are then used as the input data to the proposed MC-GAN model, and the true H&E is used as the model ground truth.

Visual assessment, and Structural Similarity Index (SSIM) [25], computed between the colorized images and their corresponding ground truth, are used to determine the optimal value of $K$. The best $K$ is selected independently for the two datasets in hand. For the human skin dataset, the $K$-study revealed that the best results are obtained when $K = 3$, whereas for the mouse brain dataset, the $K$-study SSIM results indicate an optimal value for $K$ of 2.

After extracting the features, a combinatorial study (the "$C$-study") is conducted to identify the optimal subset of features. In the $C$-study, the analysis encompasses all the available image components. By utilizing the optimal value of $K$ obtained from the $K$-study, a set of features denoted as $M_f^{opt}$ is generated. Additionally, the conventional NR and R channels are combined with $M_f^{opt}$ feature images forming an array of images $A$, where $A = [NR_{532}, NR_{266}, R_{266}, M_f^{opt}]$. This array provides a comprehensive representation of the image data. It is possible that using all the elements in $A$ to train a model could lead to redundancy and thus confuse the model. Consequently, the $C$-study conducts an exhaustive search across all the possible combinations of elements in array $A$ to determine the optimum feature combination for creating a robust colorization model. The size of array $A$ is determined by the sum of the number of features obtained from the $K$-study and the three conventional channels. Therefore, the total number of elements in array $A$ is given by $N = K + 3$. The number of possible combinations is $2^N - 1$.

To evaluate the model performance for the $C$-study, pixel-wise evaluation metrics, SSIM, Peak Signal-to-Noise Ratio (PSNR) [26], and Root Mean Squared Error (RMSE), are calculated since paired input and ground truth images are available [27]. The three metrics are computed between the colorized images and the true H&E. Images of the two domains are blurred prior to computing the metrics to avoid the effect of registration errors. Based on the assessment outcomes, a feature selection process is carried out to determine the best-performing members from set $A$, as presented in Section II.E. These are then used to train virtual staining models. The results of the two datasets are presented in the following subsections.

*Human Skin Dataset*

The number of elements in $A$ is determined based on the value of the best $K$ that is selected during the $K$-study. For instance, using $K = 3$ results in an image set $A = [NR_{532}, NR_{266}, R_{266}, m_{f_1}, m_{f_2}, m_{f_3}]$ which contains three extracted feature images along with the conventional PARS contrasts. Figure 4 (a-f) displays the members of $A$, along with their corresponding ground truth (Fig. 4 (h)). It is worth mentioning that the three extracted features do exhibit a certain level of correlation, as shown in Fig. 4 (g), however the colorization results do show that these features are less redundant and better segment the structures compared to the conventional channels.

In the $C$-study, an exhaustive search is conducted on the set $A$, as explained earlier in this section. The objective is to determine the best combination of features for colorization. Given that there are six ($K + 3$) elements in $A$ for the human skin data, there are a total of 63 possible combinations that can be used to train individual models. After training the 63 models, virtual staining is performed on unseen test data. This generates 63 colorized image sets using all the feature combinations from $A$. The test images are acquired from a distinct tissue section compared to the training data. These colorized test images are then compared against true H&E images.

Table 1 summarizes the quantitative assessment results. All of the employed metrics reached a consensus that there are at least 17 feature combinations which produce superior colorizations compared to the conventional PARS channels. Additionally, there are 13 combinations that outperform using all $N$ elements in $A$. This suggests that certain features may be redundant when combined and may potentially confuse the model learning process. Conversely, other features prove extremely valuable in colorization as they enhance the segmentation of tissue structures with distinct colors. Notably, the R channel is included in all the top 30 feature combinations, highlighting the critical role which the R contrast plays in achieving precise colorization by offering distinctive and valuable biomolecule information.

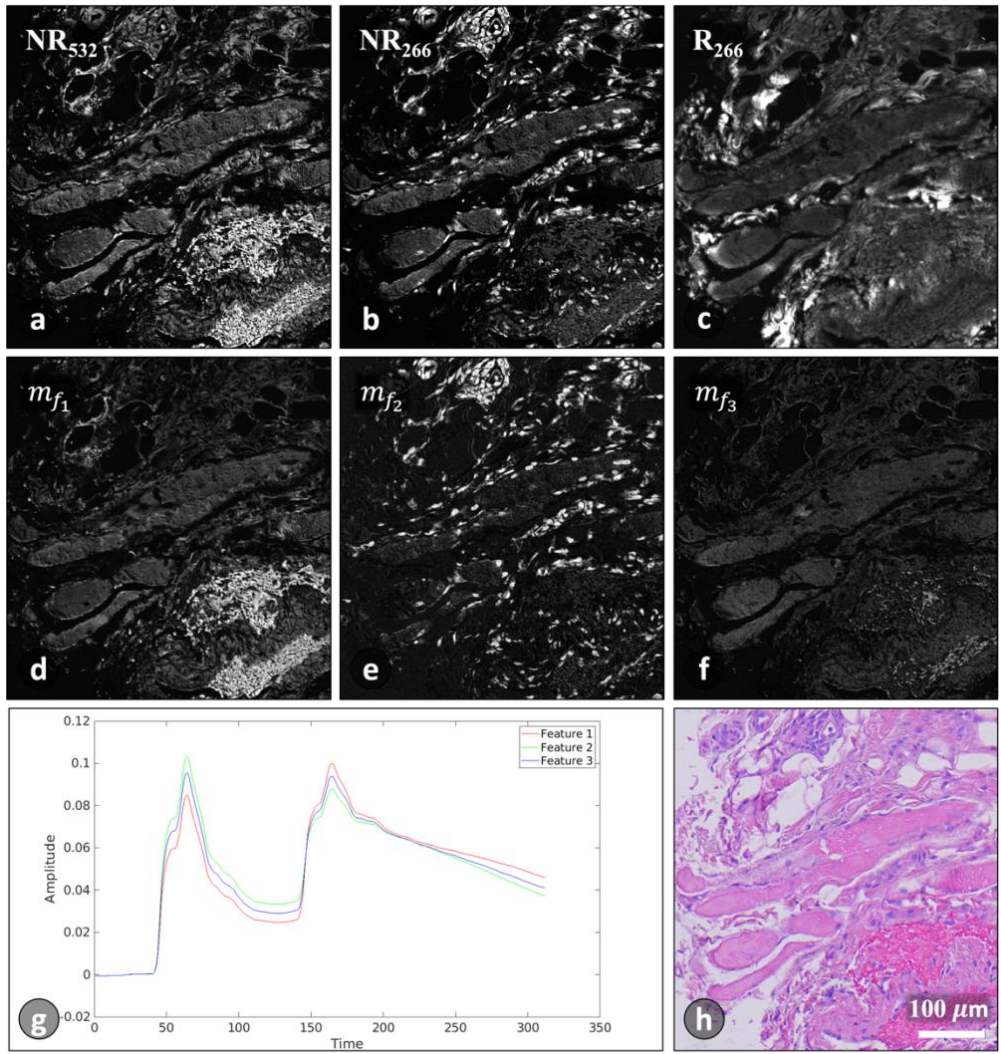

**Figure 4.** Example of PARS channels. (a) NR channel at 532 nm. (b) NR channel at 266 nm. (c) R channel at 266 nm. (d)-(f) Feature images ($m_{f_1}, m_{f_2}$ and $m_{f_3}$) corresponding to features 1-3 in (g), respectively. (g) TDs of three features extracted from NR channel. (e) True H&E of the same field-of-view.

**Table 1.** Summary of quantitative analysis of the *C*-study colorization results using the human skin dataset.

| Rank | Feature Combination | SSIM | PSNR | RMSE |
|---|---|---|---|---|
| | | | Score (rank) | |
| Best | $NR_{532}, R_{266}, m_{f_1}, m_{f_3}$ | **0.89 (1)** | 22.92 (3) | 18.22 (3) |
| | $NR_{266}, R_{266}, m_{f_1}$ | 0.88 (2) | **23.03 (1)** | **17.99 (1)** |
| | $NR_{532}, R_{266}, m_{f_2}$ | 0.88 (3) | 22.99 (2) | 18.06 (2) |
| ⋮ | | | | |
| Moderate | $NR_{532}, NR_{266}, R_{266}, m_{f_1}, m_{f_2}, m_{f_3}$ | 0.87 (14) | 22.50 (15) | 19.13 (15) |
| | $NR_{532}, NR_{266}, R_{266}$ | 0.87 (19) | 22.40 (18) | 19.35 (18) |
| ⋮ | | | | |
| Worst | $NR_{266}$ | 0.80 (61) | 19.92 (62) | 25.74 (62) |
| | $m_{f_2}$ | 0.72 (63) | 17.83 (63) | 32.71 (63) |

The colorization results are depicted in Figure 5. The least satisfactory outcomes are obtained when using only one or two features, simply too limited to capture the complexity of different structures and their corresponding stained colors. An example using only $NR_{266}$ is shown in Figure 5 (b), which represents one of the poorest three combinations. With this very limited input data, the trained model appears to mistakenly identify red blood cells as connective tissue and confuses the connective tissue with lipids, as highlighted in the yellow box. Conversely, using all the conventional channels ($NR_{266}, NR_{532}$ and $R_{266}$) in Figure 5 (c) demonstrates better performance, but still with shortcomings. For instance, some red blood cells appear more purple than they do in the ground truth H&E, as highlighted in green. In addition, the colorization of the connective tissues sometimes exhibits a mixture of purple and pink shades instead of a consistent pink tone, as shown in the blue highlighted results. These artifacts may be attributed to insufficient input information, noise, or redundancy within the input data, all of which can hinder the effective model learning.

In contrast, the feature combination which yielded the highest SSIM scores, and top three RMSE and PSNR, is shown in Figure 5 (d). In the highlighted sections, the colorization is the most accurate of the presented results. This is particularly prevalent in the red blood cells and the shading of the connective tissue. These visual comparisons between the PARS virtual H&E and the true H&E images, are supported by the quantitative measurements depicted in leftmost column of Figure 5.

The NR TD signals encompass information about the material properties being analyzed. Extracting features from these signals can significantly improve the labeling of biomolecules, leading to enhanced contrast. Notably, feature extraction surpasses the conventional image reconstruction methods in segmenting tissue structures such as nuclei, as presented in Fig. 4 (b) and (e). The effectiveness of feature extraction in tissue labelling has been previously demonstrated in studies involving PARS data of murine brain fresh tissue sample [21] and human breast tissue slide [20]. Incorporating these enhanced contrasts as input channels to a virtual staining model proves beneficial for colorization performance, ultimately enhancing the accuracy and visual representation of the colorized images.

Furthermore, the exhaustive search highlights that using an optimized set of features can be more advantageous than simply employing all available features. Targeted selection of features may lead to improved performance and reduced training time. This is likely because certain features may have a stronger ability to accurately label tissue structures, while others may be redundant and consequently cause confusion, negatively affecting the performance of the model.

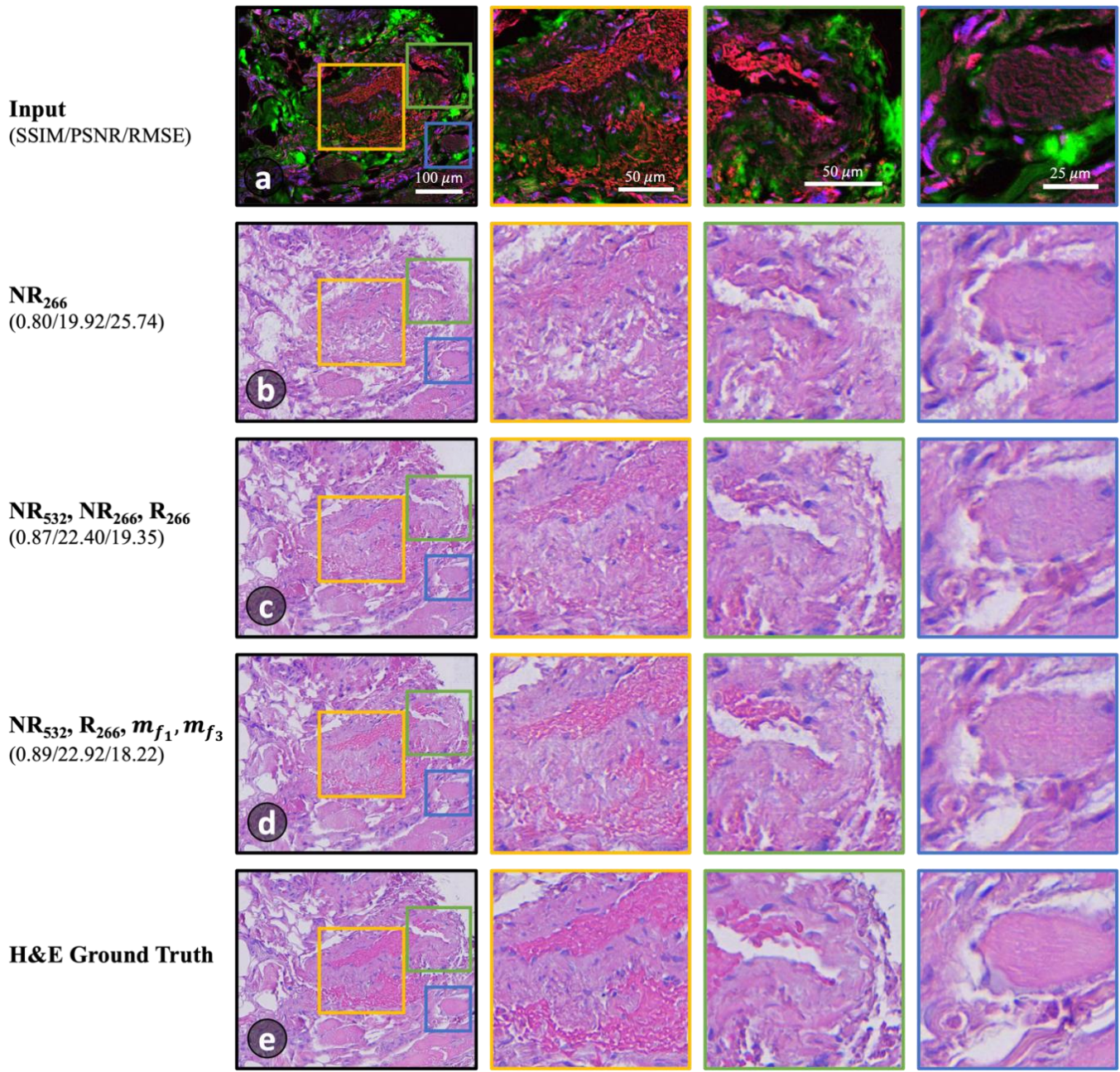

**Figure 5.** A comparison of virtual staining results using different combinations of PARS feature images as inputs. (a) RGB image of a raw PARS data where R: $NR_{532}$, G: $R_{266}$, B: $NR_{266}$ (displayed for visualization), highlighting different parts of a human skin tissue sample. (b)-(d) Worst, moderate, best results, respectively, using the feature combination labeled on the left. (e) True H&E image of the same field-of-view.

*Mouse Brain Dataset*

The mouse brain dataset shows a comparable performance pattern to the human skin. The workflow is replicated, starting with the *K*-study analysis. In this case, with an optimal K value of 2, the number of channels in domain $A$ (PARS raw data) is 5 which allows for 31 trainable models with different channel input combinations.

Following a comparison of all 31 input combinations, the quantitative assessment shows that there are at least 10 feature combinations that outperform the utilization of conventional features alone. Additionally, there are 8 alternative options that demonstrate superior performance compared to using all the features in domain $A$. Figure 6 presents a comparative analysis of the worst, moderate, and best colorization results achieved with various feature combinations as inputs to the model. The outcomes presented here further support the findings from the human dataset, which indicate that there are superior features for virtual staining that outperform the conventional features. Moreover, there is an optimal subset of features that produce best results, as opposed to all features combined.

Figure 6 (b) illustrates the outcomes obtained from a feature combination ($R_{266}$ and $m_{f_1}$) that exhibit the poorest performance among the evaluated combinations. Notably, with this input combination, the trained model erroneously classifies parts of connective tissue as cell nuclei. Upon observing Fig. 6 (b) and (c), it becomes apparent that connective tissue structures and tissue surrounding the tumor, which should ideally exhibit a distinct pink color (as demonstrated in the ground truth image, (e)), are more accurately colorized in (d) compared to (c). In (c), a significant portion of the pink color is substituted with shades of brown. The model can sometimes employ an averaging strategy for colors as a means to minimize loss [28], resulting in the prediction of colors like gray or brown when uncertain about the best color choice. In contrast, the highest level of accuracy among the presented results is observed in (d), which exhibits significant correspondence with the ground truth in terms of visual appearance and quantitative measurements. These observations provide additional evidence that the extracted features contribute to the learning process of the model. Furthermore, they emphasize the significance of selecting features that accurately label and represent the data, ultimately resulting in improved overall performance. The presented results highlight that superior feature extraction methods and feature combinations exist for improved colourization of the raw input data. Future efforts will explore other tissue types and staining varieties.

## IV. Conclusion

In conclusion, this study explores the use of the time resolved non-radiative signals for improving virtual staining of PARS images in both human skin and mouse brain. Using the K-means method [20], [21] to extract features from the non-radiative TD signals, additional information about imaged targets is captured. The proposed MC-GAN extends the conventional colorization model to accept more than three channels, allowing for the utilization of these additional features. The experimental results demonstrate that certain feature combinations outperform the conventional PARS channels as they exhibit improved labeling of tissue structures.

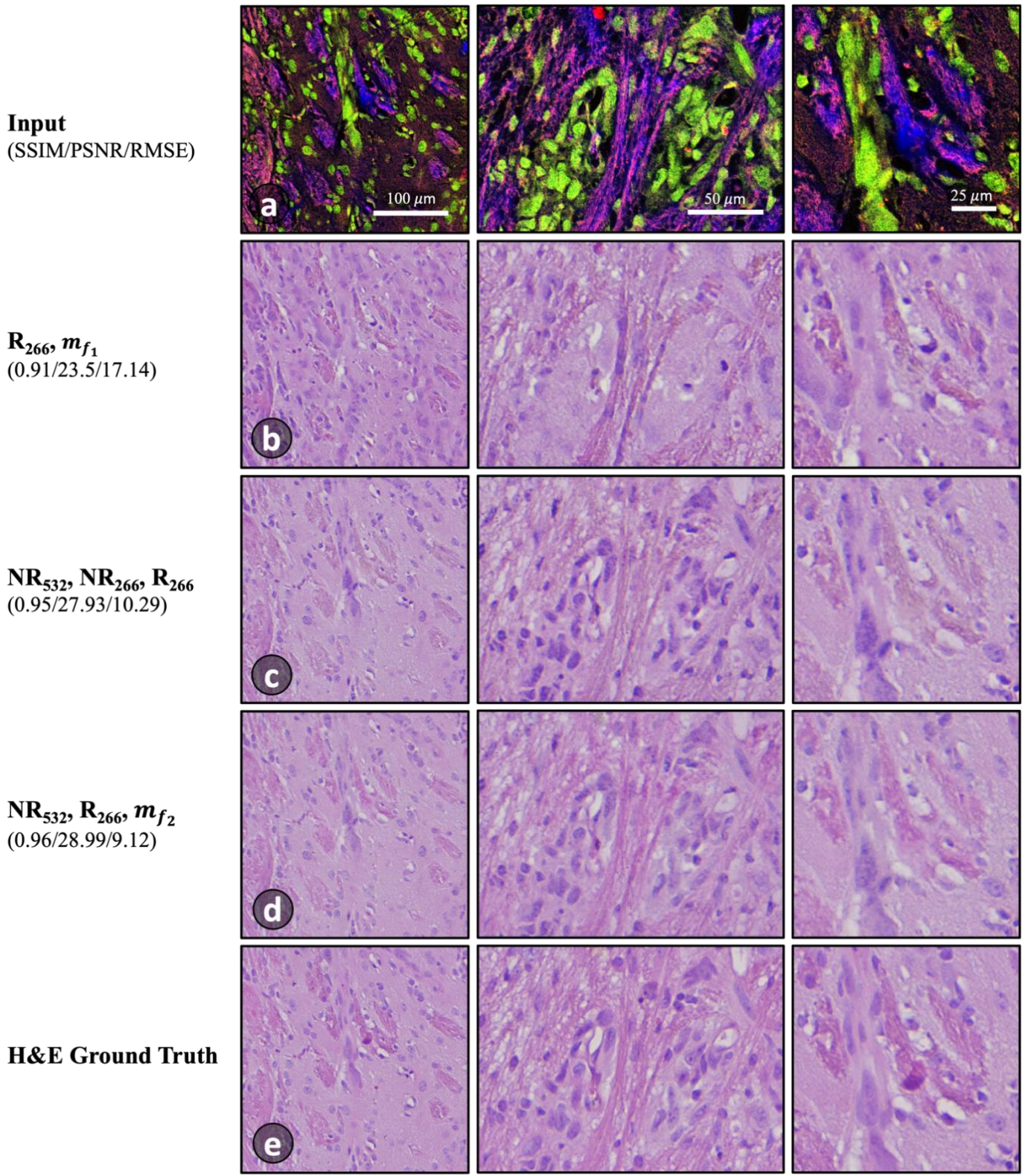

**Figure 6.** Another comparison of virtual staining results of mouse brain tissue. (a) RGB image of raw PARS data where R: $NR_{532}$, G: $R_{266}$, B: $NR_{266}$ (displayed for visualization). (b)-(d) Shows worst, moderate, and best results, respectively, using the feature combination labeled on the left. (e) True H&E image of the same field-of-view.

Several experiments are conducted to determine the optimum number of K-means features ($K$) as well as the optimum feature combination for training virtual staining models. Three different metrics are employed to evaluate the model's performance for feature selection. The limitations of using only one or two features are evident, as they fail to accurately represent the complexity of different structures and their colors. Moreover, employing all the available features can lead to confusion within the model due to the potential redundancy among them. Therefore, it was crucial to conduct a comprehensive search to identify the most effective feature combinations, which not only reduced training time by utilizing fewer features but also alleviated model confusion.

With the optimal feature combination, the colorization results exhibit a high degree of accuracy, as observed in the results. These findings highlight both high visual and quantitative agreement between the H&E-*like* PARS and the true H&E images among the two datasets, emphasizing the potential of TD signals in enhancing the accuracy of virtual staining techniques.

**Acknowledgements**

The authors thank Dr. Marie Abi Daoud at the Alberta Precision Laboratories in Calgary, Canada for providing the human skin tissue samples and Dr. Deepak Dinakaran and Dr. Kevin Camphausen from the radiation oncology branch at the National Cancer Institute, NIH, Bethesda, MD, USA for providing the mouse brain samples. Additionally, the authors would like to acknowledge Hager Gaouda for their valuable assistance in staining the tissue samples used in this study.

The authors gratefully acknowledge the financial support provided by the following funding sources throughout the duration of this project: Natural Sciences and Engineering Research Council of Canada (DGECR-2019-00143, RGPIN201906134), Canada Foundation for Innovation (JELF #38000), Mitacs Accelerate (IT13594), University of Waterloo Startup funds, Centre for Bioengineering and Biotechnology (CBB Seed fund), illumiSonics Inc (SRA #083181), New frontiers in research fund – exploration (NFRFE-2019-01012), and The Canadian Institutes of Health Research (CIHR PJT 185984).

**Author Contribution Statement**

M.B developed and implemented the multi-channel virtual staining framework, conducted experiments, prepared the figures, and wrote the main manuscript. J.E.D.T and B.R.E implemented PARS imaging system, helped with scanning PARS samples, and helped write the manuscript. J.A.Y contributed to the implementation of the cycleGAN model to support multi-channel inputs. P.F. assisted in planning the experiments and offered consultation in the writing of the manuscript. P.H.R contributed as the principal investigator, taking charge of project direction, organization, and manuscript writing.

**Competing Interests**

Authors James Tweel, Benjamin Ecclestone, and Parsin Haji Reza all have financial interests in IllumiSonics which has provided funding to the PhotoMedicine Labs. Authors Marian Boktor and Paul Fieguth do not have any competing interests.

**Data Availability**

The data that support the findings of this manuscript are available from the corresponding author, P.H.R., upon reasonable request.